\documentclass[a4paper,11pt]{article}
\usepackage{pos}

\title{Anisotropic flow decorrelation in heavy-ion collisions at RHIC-BES energies with 3D event-by-event viscous hydrodynamics}
\ShortTitle{Flow decorrelation at RHIC-BES energies with 3D event-by-event hydrodynamics}

\author*[a]{Jakub Cimerman}
\author[a]{Iurii Karpenko}
\author[a,b]{Boris Tom\'a\v{s}ik}
\author[a]{Barbara Antonina Trzeciak}

\affiliation[a]{Faculty of Nuclear Sciences and Physical Engineering, Czech Technical University in Prague,\\  B\v rehov\'a 7, 11519 Prague 1, Czech Republic}

\affiliation[b]{Univerzita Mateja Bela, Tajovsk\'eho 40, 974~01 Banská Bystrica, Slovakia}

\emailAdd{jakub.cimerman@fjfi.cvut.cz}

\def\snn{\mbox{$\sqrt{s_{_{\rm NN}}}$}}

\def\etaref{\mbox{$\eta_{\mathrm{ref}}$}}
\def\etasref{\mbox{$\eta_{s,\mathrm{ref}}$}}
\def\trento{T$_{\rm R}$ENTo}

\abstract{In the RHIC Beam Energy Scan program, gold nuclei are collided with different collision energies in the range from few to 62.4 GeV. The goals of the program are to explore the onset of QGP creation, locate the critical point of QCD and study dense baryon matter.
We report on the first application of Monte Carlo Glauber (GLISSANDO2) and \trento~$p=0$ initial states extended to 3D for event-by-event viscous fluid dynamic (vHLLE) with hadronic cascade modelling of Au+Au collisions at $\snn=27$ and 62.4 GeV, which is the upper region of RHIC BES energies. The initial states are extended into both the longitudinal direction and for finite baryon density using simple ansätze. The full energy and baryon charge counting in the initial states is implemented. We show the reproduction of elliptic flow, at both collision energies and with both initial states. We compare it also to the results obtained with UrQMD initial state.
Furthermore, we show the results for rapidity decorrelation of elliptic flow $r_2$ at $\snn=27$ and 200 GeV from the same setup of hydrodynamic calculations with the 3D Monte Carlo Glauber and UrQMD initial states. We discuss the features of the initial states responsible for the magnitude of the observed flow decorrelation.}

\FullConference{%
  *** The European Physical Society Conference on High Energy Physics (EPS-HEP2021), ***\\
  *** 26-30 July 2021 ***\\
  *** Online conference, jointly organized by Universität Hamburg and the research center DESY ***
}

\begin{document}
\maketitle

\section{Introduction}

For two decades, investigations of
anisotropic flows taught us much about the quark-gluon plasma, but most of
the studies focused on the flow in the transverse plane at midrapidity. However, studying event-by-event fluctuations along the longitudinal direction may help us understand
the transport properties of quark-gluon plasma.
At RHIC-BES energies, the decorrelation of the flow anisotropy along the longitudinal direction is just starting to be researched. So far, there are only preliminary results from STAR at $\snn = 27$
and 200~GeV \cite{Nie:2019bgd,Nie:2020trj}. This paper summarizes the first calculation of a kind at RHIC-BES energy
in a hydrodynamic model \cite{Cimerman:2021gwf}.

\section{Model}

For the calculations a hybrid event-by-event viscous hydrodynamic model is used. It consists of four parts. It starts with three-dimensional initial state. Here, we worked with three different models: UrQMD \cite{Bass:1998ca}, which uses PYTHIA6 to simulate inelastic nucleon-nucleon scatterings through string formation and subsequent string break-up, GLISSANDO2 \cite{Rybczynski:2013yba}, which is a Monte Carlo Glauber model, and \trento ~\cite{Moreland:2014oya}, which introduces a generalized ansatz for the entropy density deposition from the participant nucleons. Since Glissando and \trento ~only provide the initial state in the transverse
plane, we extended them to longitudinal direction following \cite{Bozek:2012fw}.
Moreover, we fixed the total energy and baryon charge to those of
participants and therefore there is one less parameter to tune. The hydrodynamic
stage of evolution is modelled with a hydrodynamic code vHLLE \cite{Karpenko:2013wva}, in which we included the shear viscosity. Next,
a Monte Carlo hadron sampling is performed according to Cooper-Frye formula.
The final step of the model is to simulate hadronic rescatterings and resonance
decays of the sampled hadrons using UrQMD cascade. The model also includes
a finite baryon and electric charge density at all stages and it is important,
because we are doing simulations at relatively low energies. More detailed description of the model can be found in \cite{Cimerman:2020iny}.

\section{Flow}

\begin{figure}[h]
    \centering
    \includegraphics[width=0.48\textwidth]{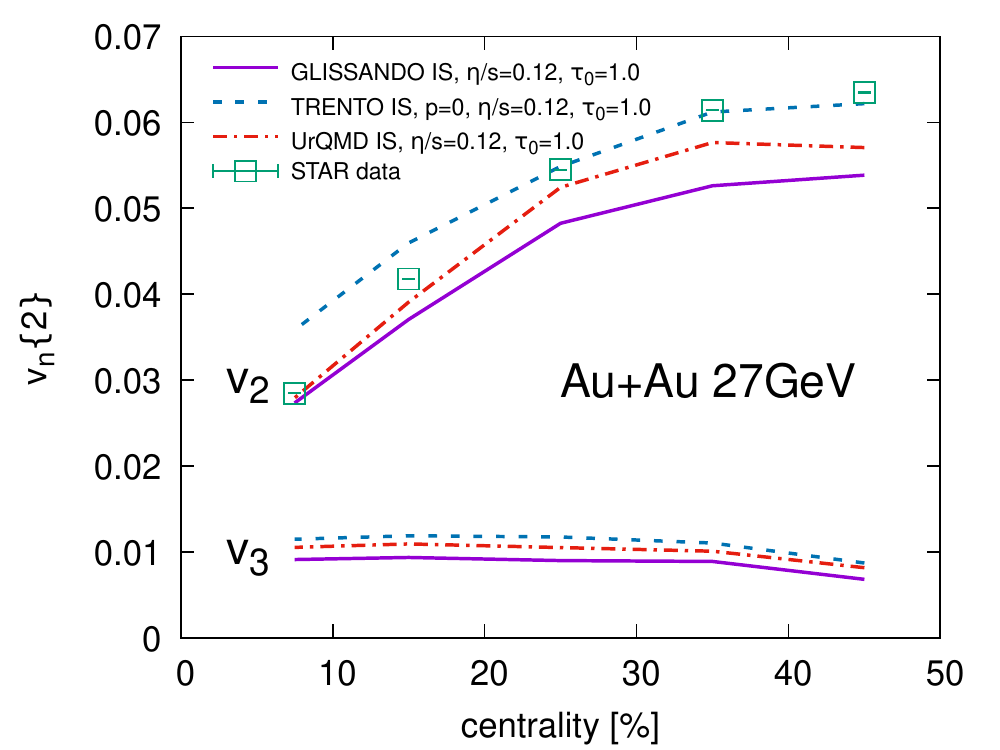}
    \includegraphics[width=0.48\textwidth]{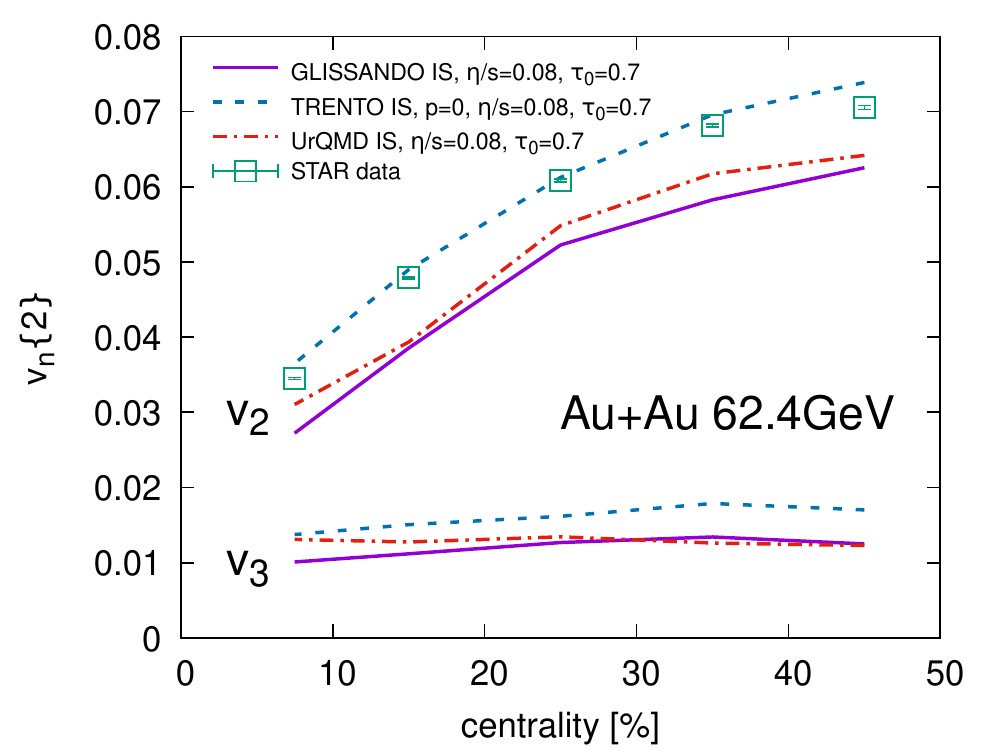}
    \caption{Elliptic and triangular flows of charged hadrons as functions of centrality for $\snn=27$ (left) and 62.4 GeV (right) calculated using 2-particle cumulant method compared with STAR data \cite{STAR:2012och,STAR:2017idk}.}
    \label{fig:v2-centrality}
\end{figure}

\begin{figure}[h]
    \centering
    \includegraphics[width=0.48\textwidth]{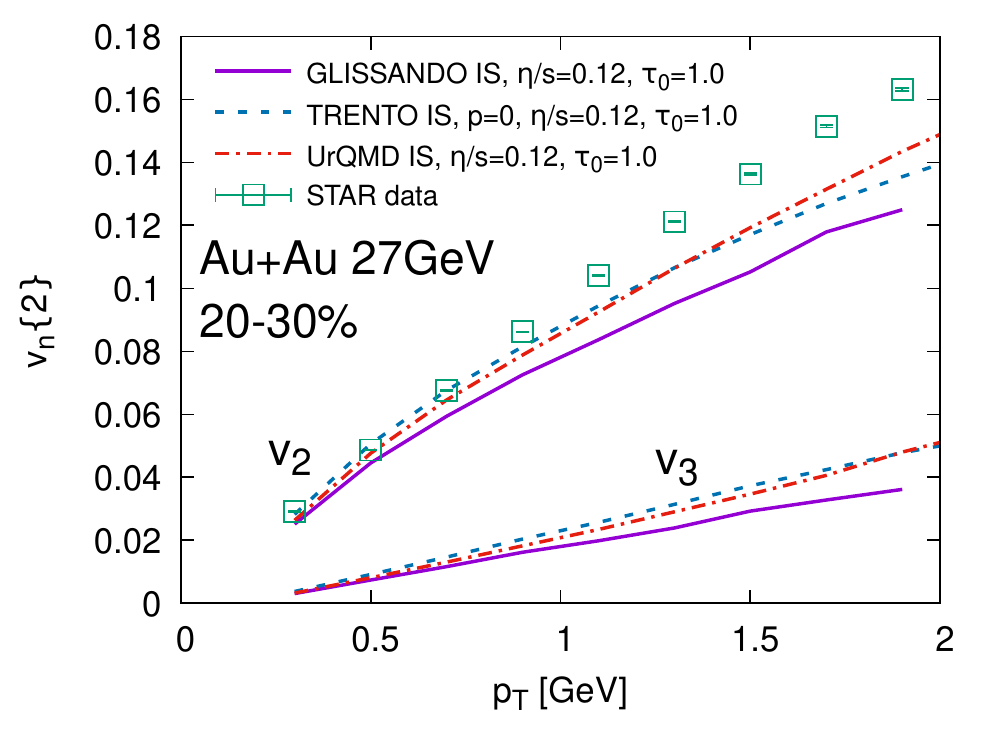}
    \includegraphics[width=0.48\textwidth]{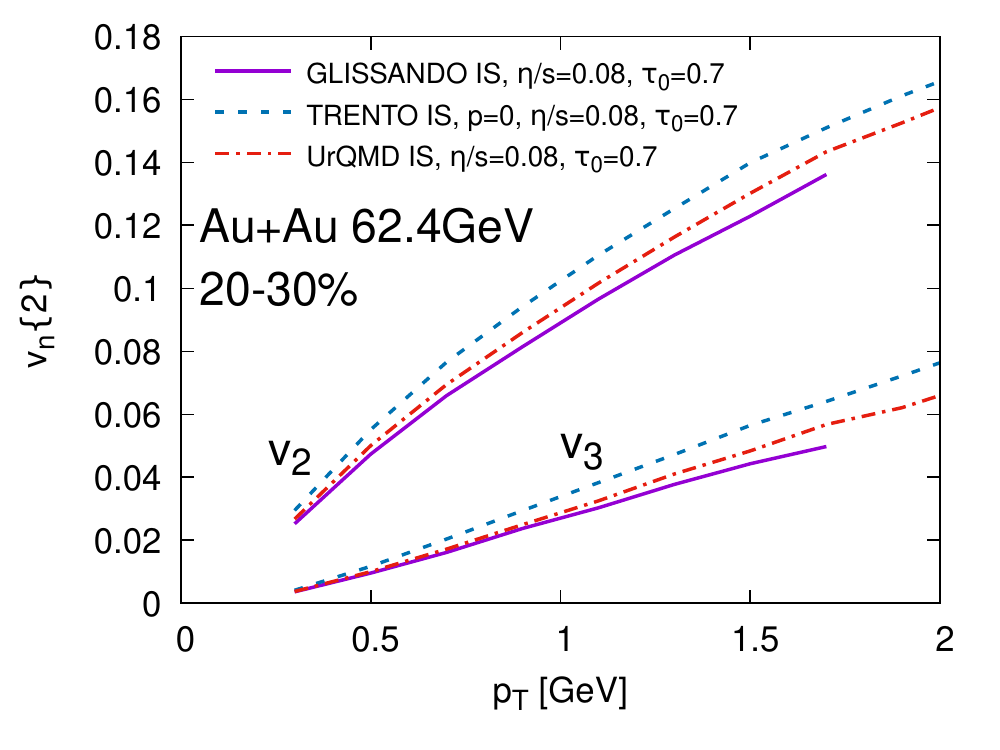}
    \caption{Elliptic and triangular flows of charged hadrons as functions of $p_T$ for $\snn=27$ (left) and 62.4 GeV (right) calculated using 2-particle cumulant method compared with STAR data \cite{STAR:2012och}.}
    \label{fig:v2-pt}
\end{figure}

The classical definition of the anisotropic flow is through a Fourier series of $p_T$ distribution:
\begin{equation}
    v_n=\dfrac{\int \mathrm{d}\phi \cos(n(\phi-\Psi_n))\frac{\mathrm{d}^3N}{p_T\mathrm{d}p_T\mathrm{d}y\mathrm{d}\phi}}{\int \mathrm{d}\phi\frac{\mathrm{d}^3N}{p_T\mathrm{d}p_T\mathrm{d}y\mathrm{d}\phi}}.
\end{equation}
There are several methods to calculate anisotropic flow from experimental or simulated data, from which the event plane (EP) method and the 2-particle cumulant method were used for this work. Figure \ref{fig:v2-centrality} shows the centrality dependence of the $p_T$ integrated elliptic and triangular flow. From this figure we can see that \trento~IS produces the largest anisotropic flow, whereas GLISSANDO IS produces the lowest $v_2$ and $v_3$. In the non-central collisions
at 27~GeV as well as in all considered centrality classes at 62.4~GeV, calculations
with \trento~IS have the best agreement with the experimental data for elliptic
flow. However, $p_T$ dependent elliptic flow (Figure \ref{fig:v2-pt}) shows that our calculations are in agreement with the data only for $p_T<1$~GeV. At larger $p_T$, all of the initial states start to under-predict the data.

\begin{figure}[h]
    \centering
    \includegraphics[width=0.48\textwidth]{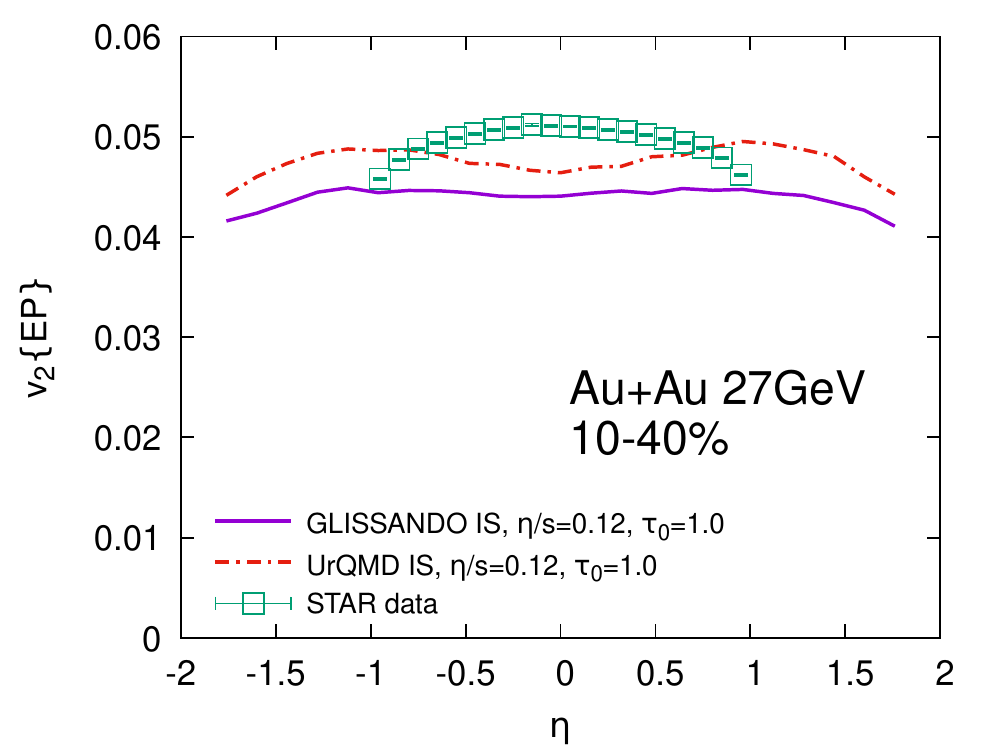}
    \includegraphics[width=0.48\textwidth]{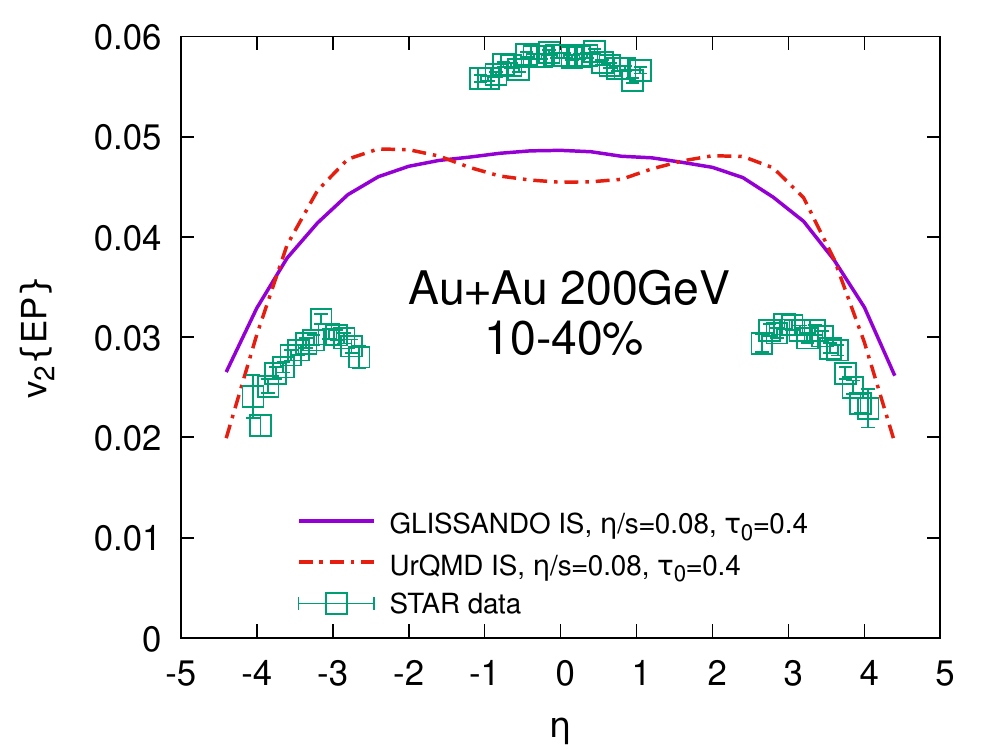}
    \caption{Elliptic flow of charged hadrons as a function of pseudorapidity for $\snn=27$ (left) and 200 GeV (right) calculated using EP method compared with STAR data \cite{STAR:2012och,STAR:2008ftz}.} 
    \label{fig:v2-eta}
\end{figure}

Figure \ref{fig:v2-eta} shows the elliptic flow as a function of pseudorapidity. At 27~GeV, GLISSANDO and UrQMD initial states reproduce overall order
of magnitude of the elliptic flow, but underestimate its value at mid-rapidity,
which is consistent with previous figures. At 200~GeV the experimental data indicate a triangular pseudorapidity dependence that neither of the initial states
can describe.

\section{Decorrelation}

The longitudinal fluctuations leads to decorrelation of anisotropic flows along the pseudorapidity direction. Using the definition of the flow vector $\textbf{V}_n=v_n e^{in\Psi_n}$, the longitudinal decorrelation can be expressed using the factorization ratio:
\begin{equation}
    r_n(\eta, \etaref)=\dfrac{\left\langle\textbf{V}_n(-\eta)\textbf{V}^\ast_n(\etaref)\right\rangle}{\left\langle\textbf{V}_n(+\eta)\textbf{V}^\ast_n(\etaref)\right\rangle}=\dfrac{\left\langle v_n(-\eta)v_n(\etaref)\cos n(\Psi_n(-\eta)-\Psi_n(\etaref))\right\rangle}{\left\langle v_n(+\eta)v_n(\etaref)\cos n(\Psi_n(+\eta)-\Psi_n(\etaref))\right\rangle}.
    \label{eq:r2}
\end{equation}
If the factorisation ratio is equal to 1, it means, that the flow vectors at
$+\eta$ and $-\eta$ are fully correlated, and when it gets below 1, we got
a decorrelation between flow vectors at $\pm\eta$. From Eq. \eqref{eq:r2} it can be 
seen that the flow decorrelation may be caused by two separate effects: flow
magnitude decorrelation and flow angle decorrelation. Thus we can define corresponding factorisation ratios:
\begin{subequations}
\begin{align}
    r_n^v(\eta)&=\dfrac{\left\langle v_n(-\eta)v_n(\etaref)\right\rangle}{\left\langle v_n(+\eta)v_n(\etaref)\right\rangle}, \\
    r_n^\Psi(\eta)&=\dfrac{\left\langle \cos n(\Psi_n(-\eta)-\Psi_n(\etaref))\right\rangle}{\left\langle \cos n(\Psi_n(+\eta)-\Psi_n(\etaref))\right\rangle}.
\end{align}
\end{subequations}

\begin{figure}[h]
    \centering
    \includegraphics[width=0.48\textwidth]{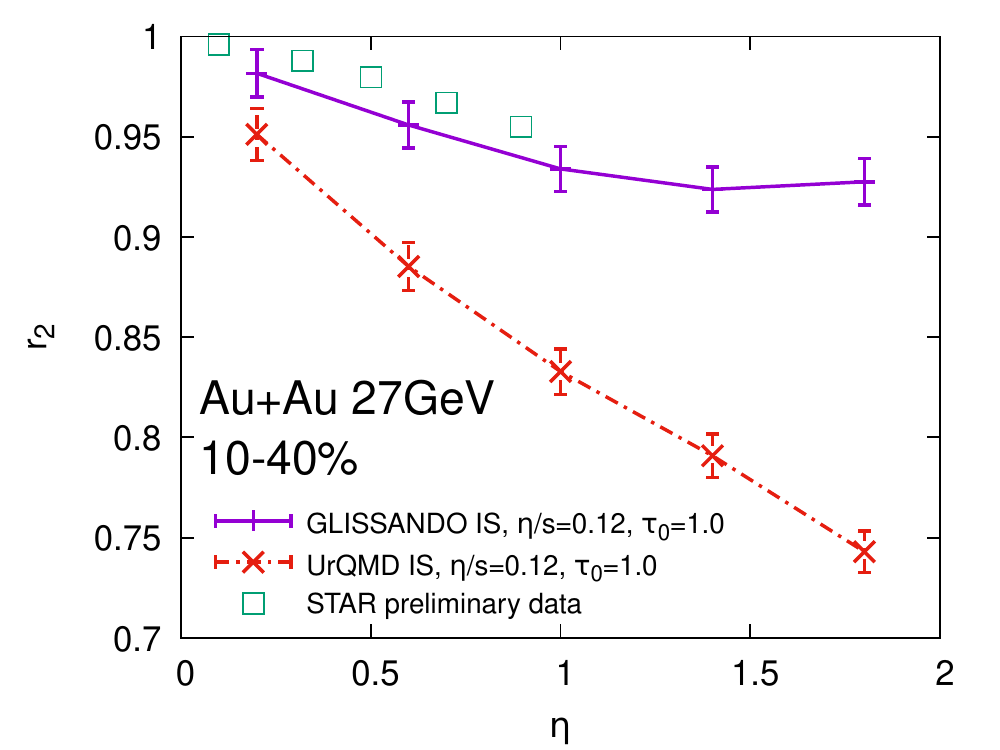}
    \includegraphics[width=0.48\textwidth]{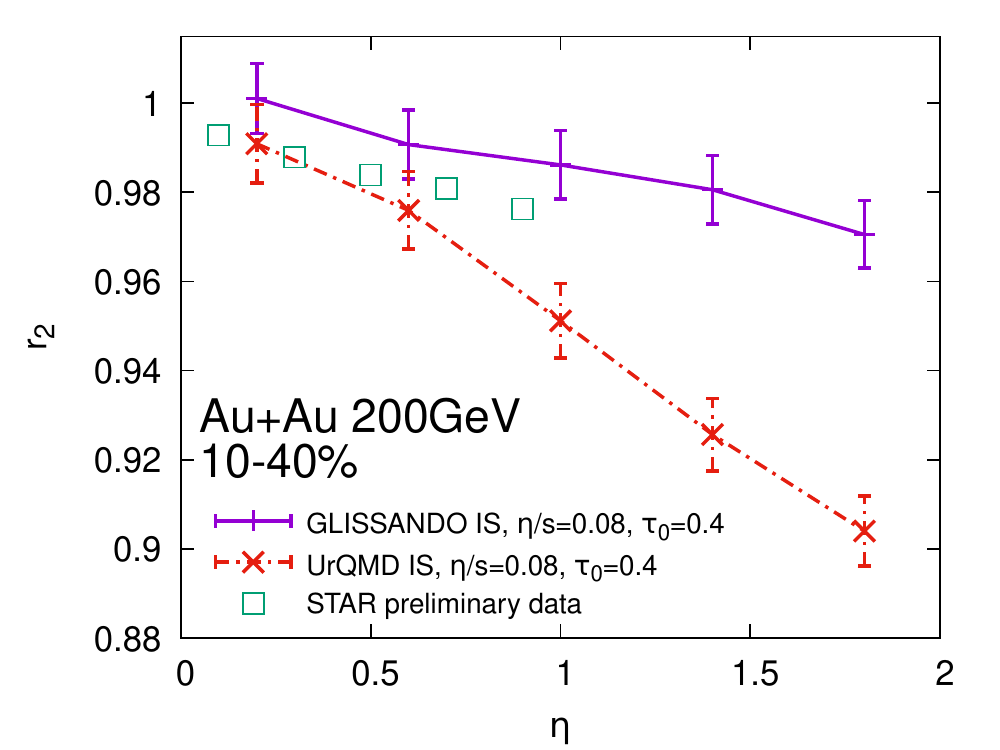}
    \caption{Factorization ratio $r_2$ of charged hadrons as a function of pseudorapidity for $10-40\%$ Au-Au collisions at $\snn=$ 27 (left) and 200 GeV (right) compared with STAR preliminary data \cite{Nie:2019bgd,Nie:2020trj}.}
    \label{fig:r2}
\end{figure}

Figure \ref{fig:r2} shows the factorisation ratio as a function of pseudorapdity. We did not use the extended \trento~IS for these calculations, since it does not have implemented any tilt in the longitudinal direction and thus the factorisation ratio would be equal to 1. At 27~GeV, the model with UrQMD initial state shows much stronger decorrelation
then the one seen in the data. On the other hand, calculations with GLISSANDO
initial state can describe the experimental data within uncertainties for both
energies.

\begin{figure}[h]
    \centering
    \includegraphics[width=0.48\textwidth]{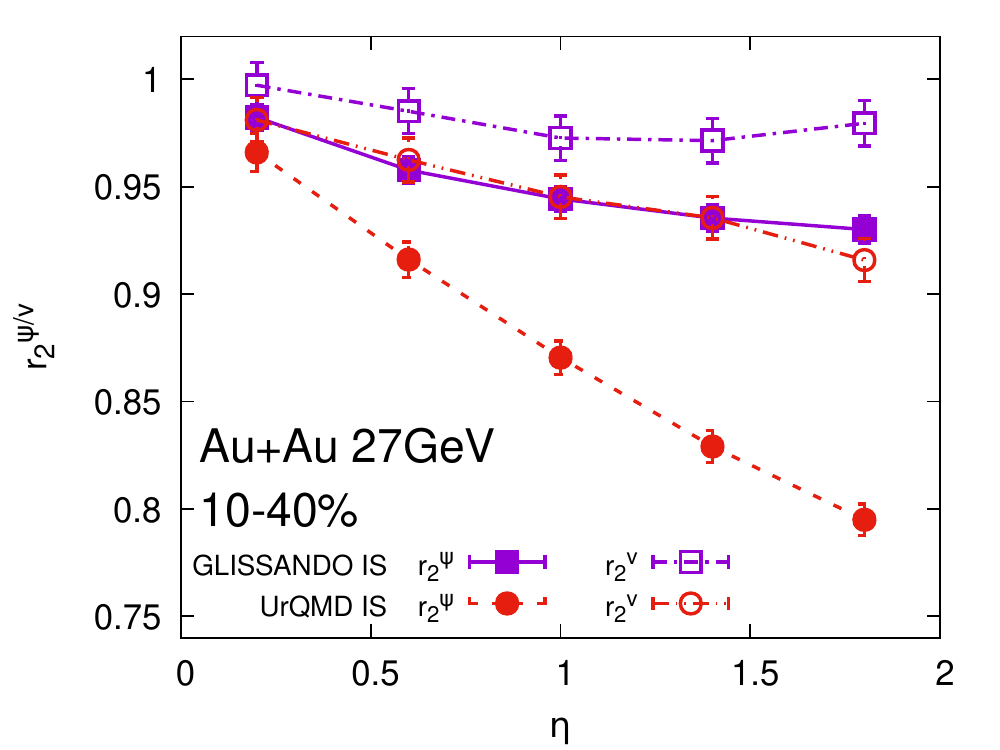}
    \caption{The flow magnitude $r^v_2$ and the flow angle decorrelation $r^\Psi_2$ of charged hadrons as a function of pseudorapidity for $10-40\%$ Au-Au collisions at $\snn=$ 27 GeV.}
    \label{fig:r2_v_psi}
\end{figure}

Figure \ref{fig:r2_v_psi} shows contributions of the flow angle and flow magnitude decorrelation separately for 27~GeV. It can be seen that
for both models the flow angle decorrelation plays more important role than
the flow magnitude decorrelation. The same hierarchy has been observed in
calculations at LHC energies \cite{Bozek:2017qir}.

\begin{figure}[h]
    \centering
    \includegraphics[width=0.48\textwidth]{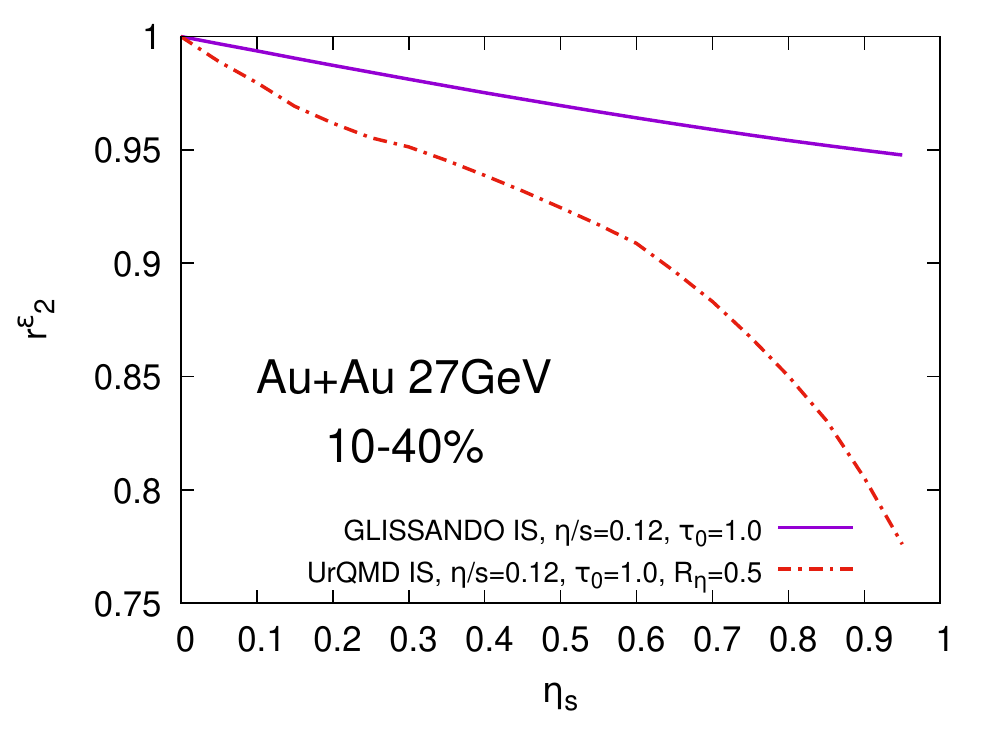}
    \includegraphics[width=0.48\textwidth]{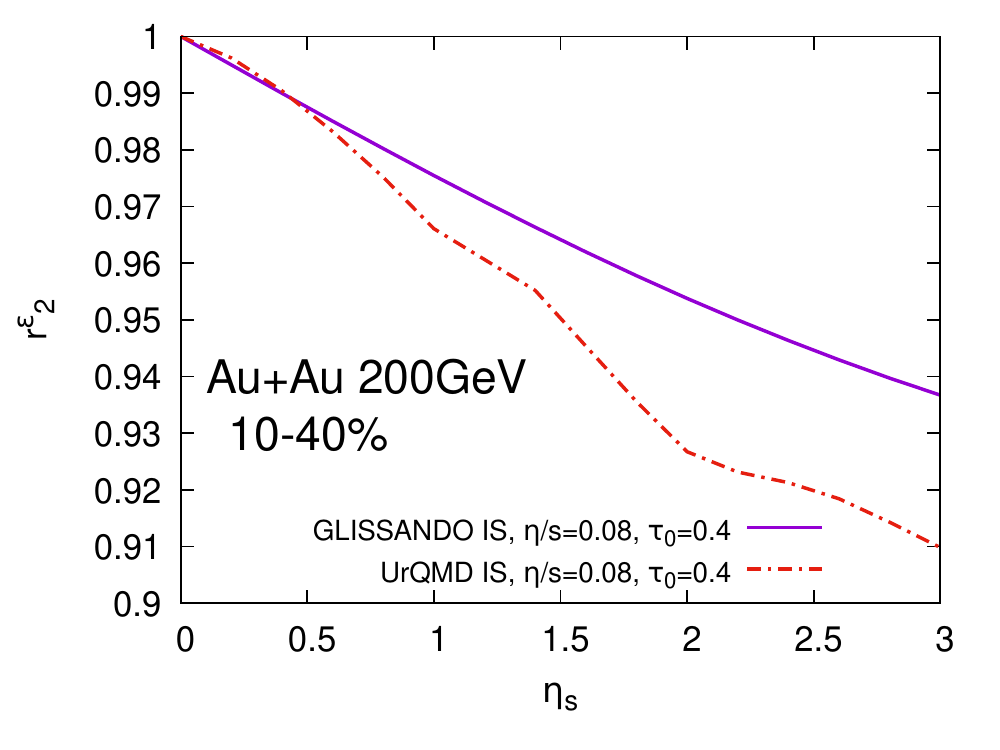}
    \caption{Longitudinal decorrelation of the initial state eccentricity $\epsilon_2$ as a function of space-time rapidity for $10-40\%$ Au-Au collisions at $\snn=$ 27 (left) and 200 GeV (right).}
    \label{fig:eps_decorrelation}
\end{figure}

It is well known that the anisotropic flow coefficients are strongly correlated
with the initial state eccentricities. Therefore, to better understand the big
difference between the initial state models, we can define the factorisation ratio
of initial-space eccentricities:
\begin{equation}
    r^\epsilon_n(\eta_s)=\frac{\left\langle \epsilon_n(-\eta_s)\epsilon_n(\etasref) \cos[n\left(\Psi_n(-\eta_s)-\Psi_n(\etasref)\right)]\right\rangle}{\left\langle \epsilon_n(\eta_s)\epsilon_n(\etasref) \cos[n\left(\Psi_n(\eta_s)-\Psi_n(\etasref)\right)]\right\rangle}
\end{equation}
where
\begin{equation*}
    \epsilon_n e^{in\Psi_n}=\frac{\int e^{in\phi}r^n\rho(\vec{r}) d\phi\, r \,dr}{\int r^n \rho(\vec{r}) d\phi\, r \,dr}
\end{equation*}
This factorisation ratio as a function of space-time
rapidity is shown in Figure \ref{fig:eps_decorrelation}. Comparison with the
flow decorrelation (Figure \ref{fig:r2}) shows that the two factorisation ratios agree even quantitatively, which means that longitudinal decorrelation is created in the initial state.

\section{Conclusions}

This paper presents the elliptic flow and the
flow decorrelation in Au-Au collisions at RHIC-BES energies simulated with the help of 3D viscous hydrodynamic model with three different initial state alternatives, which is a first simulations with this kind of hydrodynamic model at these energies. The best description of
the data at midrapidity has been provided by the model with \trento~initial state with parameter $p=0$. We found that the flow decorrelation is mainly caused by
the decorrelation of the flow angle, which has been seen also at LHC energies.
We also showed that the strong decorrelation of the model with UrQMD IS
is caused by strong decorrelation of initial-state eccentricity.

\subsection*{Acknowledgements}
JC, IK, and BT acknowledge support by the project Centre of Advanced Applied Sciences, No.~CZ.02.1.01/0.0/0.0/16-019/0000778,  co-financed by the European Union. JC and BAT acknowledge support from from The Czech Science Foundation, grant number: GJ20-16256Y. IK acknwowledges support by the Ministry of Education, Youth and Sports of the Czech Republic under grant ``International Mobility of Researchers – MSCA IF IV at CTU in Prague'' No.\ CZ.02.2.69/0.0/0.0/20\_079/0017983. BT acknowledges support from VEGA 1/0348/18. Computational resources were supplied by the project ``e-Infrastruktura CZ'' (e-INFRA LM2018140) provided within the program Projects of Large Research, Development and Innovations Infrastructures.

\end{document}